# Efficiency of a smoke curtain in a ventilated tunnel

*A Narcisse, O Vauquelin, E Casalé*
Aix-Marseille Université, Département de Mécanique, Marseille

*R Nottet*
FluidAlp, Marseille

**ABSTRACT**

Smoke curtains are typically used in public-access buildings in connexion with ventilation effects without considering a crossed design.
This paper aims to understand the fire smoke behaviour in the context of the interaction between mechanical ventilation and smoke curtains. This interaction is analysed here in the configuration of a tunnel with using numerical simulations.

Initially, without the presence of fire smoke, the length of the vortex induced downstream of the curtain is determined as a function of the longitudinal velocity and the size of the curtain. It can be observed that for a sufficiently high velocity (i.e. a sufficiently high Reynolds number), the size of this vortex depends only on the height of the smoke curtain.

Then, in the presence of a moderate fire heat release rate, we determined the air velocity required to prevent smoke from rising beyond a curtain measuring 1/5 the height of the tunnel. A significant reduction in this longitudinal velocity was observed in comparison to the velocity required to achieve the same level of containment without the presence of a curtain. The vortex generated by the curtain nevertheless interacts with the smoke layer, locally increasing its thickness.

Lastly, this configuration is tested in the case of a medium and high fire heat release rate in a road tunnel with transverse ventilation. The work carried out suggests that the installation of smoke curtains of an appropriate size, combined with control of ventilation effects, is likely to reduce the need for them.

**NOMENCLATURE**

H : tunnel height (m)
h : curtain height (m)
$h_t$ : thickness of the vortex downstream of the curtain (m)
L : distance between the fire source and the curtain (m)
$l_t$ : length of the vortex downstream of the curtain (m)

P : heat release rate of the fire (W)
$U_\infty$: longitudinal ventilation velocity (m/s)

# 1  INTRODUCTION

In confined spaces, in a fire situation, performance-based control of smoke movement is traditionally based on two very different techniques:
- Volume partitioning and forced smoke confinement. This approach is adopted in environments where a fire is likely to be the source of potentially catastrophic events (industry, nuclear installations).
- Active control using mechanical ventilation systems (road tunnels, metro systems).

Several tools can be used to contain smoke. For example, the air curtain technique has been studied in the past to determine the most suitable air flows and opening angles for smoke containment in case of a tunnel fire. Gao et al. (1) experimentally studied the effectiveness of such a solution. Nevertheless, its efficiency is limited to low heat release rate.

Confinement can also be achieved using solid smoke curtains. They obstruct the upper part of the volume to protect or the full section. These curtains are already used in many public spaces. Flexible curtains covering the entire tunnel section have already been designed. They obviously allow the smoke to be contained. Nevertheless, such a configuration (dead-end tunnel) differs from the geometry investigated here.

In this article, we propose to study a hybrid technique based on the use of solid smoke curtains. These devices are commonly used to control the smoke behaviour in public-access buildings in case of fire. They are usually designed from a passive point of view, i.e. without considering the natural or mechanical ventilation.

The technique based on smoke curtains combined with an adapted ventilation configuration appears to be an intermediate approach, since the objective (smoke confinement) uses both the physical means (smoke curtains or, more generally, geometric singularities) and the management of the aeraulic context (regulation of mechanical ventilation). This approach can appear very restrictive since it requires the implementation of major parts of each of the two techniques. Moreover, the interactions between ventilation systems, smoke layers, and smoke curtains are not well-documented, making it challenging to predict and manage their combined effects.

Although such a configuration has received little attention in the literature, we can nonetheless mention three articles that have taken an interest in the subject. The first is by Bettelini (2) who, in 2012, carried out numerical simulations with the software FDS to study the use of flexible curtains in a longitudinally ventilated tunnel for a 30 MW fire. He has shown that their use could reduce the propagation distance of the smoke layer upstream of the fire, even with a low longitudinal ventilation of around 1 m/s. His simulations have also shown that the use of curtains increases the time taken for smoke to propagate upstream of the fire location, which appears obvious.

Based on small-scale experiments, Chaabat et *al*. (3) investigated the effect of a smoke curtain placed immediately upstream of a fire source, on reducing the critical velocity in a longitudinally ventilated tunnel. The authors have shown that it was possible to reduce the longitudinal velocity value by 10% to 40%, depending on the height of the curtain.

However, their results have a rather limited applicative scope since the curtain was placed close to the source.

More recently, Halawa & Saftwat (4) and Halawa (5) have extended the numerical work of Bettellini (2) by considering larger and more complex curtains. Their results show how the association of many curtains can be beneficial, in all cases, to prevent smoke propagation against ventilation.

The present paper completes the previous works by investigating the velocity required to confine smoke behind a curtain. Containment velocities with and without curtains will be compared to highlight the potential gains from using curtains in tunnels. This containment velocity with a curtain will be compared with the confinement velocity for which the layer stabilises at the same distance upstream from the fire but without curtains.

At first, in sections 2, 3 and 4, we consider a longitudinally ventilated tunnel where a curtain is placed upstream of the fire source in order to stop the backlayering progression against the main flow. We propose here to numerically investigate a simple configuration for a 2 MW fire with a single curtain placed at a significant distance L of the order of 10 times the tunnel height upstream from the fire location. The smoke curtain is initially studied without fire, to identify the generated aerodynamic disturbances. Simulations with a fire are then carried out, first without a curtain with a longitudinal airflow that stabilises backlayering at the distance L. With the presence of a curtain (one-fifth the height of the tunnel), the longitudinal velocity is progressively decreased until the smoke layer spills out beyond the curtain.

Then, in section 5, we consider a road tunnel with 1m high containment curtains and a transverse ventilation system. The tunnel is 5m high and 10m wide. Numerical simulations are carried out for three different fire heat release rates: 2, 5 and 10 MW. In this case, the extraction flow rate is adjusted (and not the longitudinal velocity) in order the smoke to be contained between the curtains. The longitudinal velocity associated to this extraction flow rate is called the confinement velocity (6).

## 2  GEOMETRY AND NUMERICAL PROCEDURES

The configuration under consideration in the first part refers to a 200-m-long tunnel which cross-section is 5 m (height) x 10 m (width). A velocity $U_\infty$ is set at the tunnel entrance. The tunnel exit is open to external pressure. In simulations involving a fire, the source is located 100 m from the tunnel entrance, i.e. in the middle of the tunnel and centred in the cross section. The heat release rate is fixed at 2 MW (horizontal surface: 1 m²). No heat release growth is considered, and that the permanent maximum combustion is reached instantaneously.

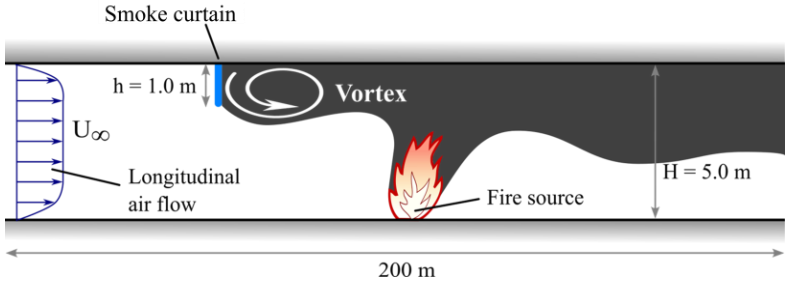

Figure 1: Geometric configuration of the academic model.

In configurations involving a smoke curtain this will be placed upstream of the fire source at a distance to be discussed in section 3. Its height, however, will be kept constant at 1 m.

About academic study, simulations were conducted using the FDS (7) version 6.9.1 (Fire Dynamic Simulator) calculation code developed by NIST (National Institute of Standards and Technology). The mesh is structured comprising about $1.3.10^6$ cubic meshes of 25.0 cm and 12.5 cm sides. The grid is refined in the vicinity of the walls and the fire source. Each numerical test simulates a 10 min period. The chronology is as follows:

- $t = 0$ s : the velocity is imposed at the domain entrance and the flow develops during 200 s.
- $t = 200$ s : the fire is activated (2 MW) and the flow develops during 300 s.
- $t = 500$ s : start data acquisition (velocity and temperature).
- $t = 600$ s : end of the simulation.

The average fields (velocity and temperature) are therefore calculated between $t = 500$ s and $t = 600$ s.

## 3      AERODYNAMIC DISTURBANCES DOWNSTREAM OF A CURTAIN

Analysis of the flow pattern relevant to the aerodynamic disturbances generated downstream of the curtain provides a clear description of the aeraulic context and a better understanding of the phenomena developed by a fire in this situation.

For an average longitudinal velocity of 1 m/s and a 1 m, the streamlines obtained are shown in Figure 2 below.

We note the development of a stable recirculation zone which length is approximately about 11 to 12 times the height of the curtain. This result is in accordance with those reported in the literature, and with experimental data obtained by Bergeles and Athanassiadis (8) for similar values of h/H ratios.

This raises the question of the effect of this recirculation zone on the stability of the backlayering and more generally on the interaction between the two structures. They are to be considered as two independent vortices with identical rotation directions.

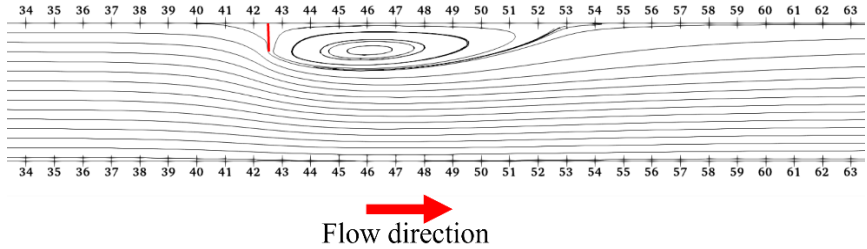

**Figure 2: Air flow (streamlines) in a tunnel with a smoke curtain.**

Beyond the recirculation zone, the presence of a curtain will locally increase the longitudinal velocity through contraction of the flow cross-section. This acceleration may affect the stratification of the backlayering and the fire thermal plume. The simulations presented in the following section aim to provide answers to these questions.

## 4    EFFECTIVENESS OF THE SMOKE CURTAIN ON THE SMOKE CONFINEMENT

The simulations are carried out in a fixed configuration with a longitudinal air velocity of 1 m/s and a fire heat release rate of 2 MW. The first simulation, carried out without a smoke curtain, results in the stabilization of backlayering propagation at 57.5 m from the fire source. The second simulation is performed with a smoke curtain located at this distance.
Figure 3 below present a comparison of the iso-temperature contours obtained for the two simulations: (a) without a curtain and (b) with a curtain.

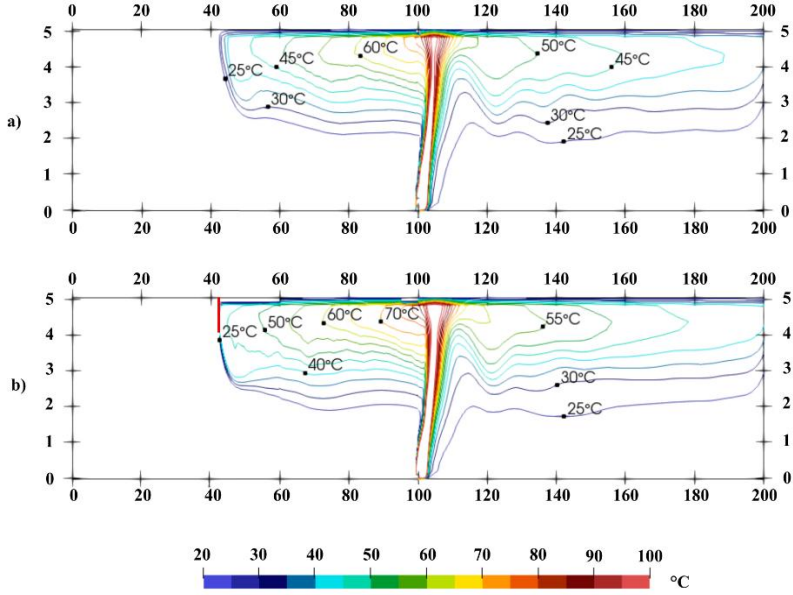

**Figure 3: Isothermal lines obtained for a 2 MW fire and a longitudinal velocity of 1 m/s: (a) without a curtain and (b) with a curtain.**

In the case with a curtain, the smoke is still confined, but there is a greater thickness, which can be attributed to the effect of the recirculation zone downstream of the curtain. The fire plume does not appear to be tilted, despite the increase in longitudinal velocity beneath the smoke curtain. The longitudinal velocity profile at the fire location appears to have been restored. The isotherms seem to show a slight increase in temperature in the case with the curtain, which could reinforce stratification in the flow downstream of the focus. To quantify this stratification, we propose to calculate the Newman number (8) defined by:

$$C_N = \frac{T_{ceiling} - T_{floor}}{T_{average} - T_{ambient}} \tag{1}$$

About 50 m downstream from the fire source, $C_N$ values close to 2.0 are obtained in both cases (with and without a smoke curtain) which, according to Newman, corresponds to a rather stratified situation (the destratification relates to $C_N < 1.7$). Logically, there is no noticeable increase in downstream stratification with the use of a smoke curtain.

If we refer to the conclusions of the article by Chaabat et al. (3), the main advantage of a curtain is to be able to reduce the longitudinal ventilation velocity to ensure smoke control. We therefore carried out additional simulations with progressively lower values of the longitudinal velocity until the curtain could no longer ensure smoke blockage. The results are presented n Table 1, in which, for different values of longitudinal velocity, the maximum thickness of the layer between the smoke curtain and the fire and the Newman number at 50 m downstream of the fire are given.

Table 1 : Simulation results

| Velocity [m/s] | Smoke overflow at the smoke curtain | Newman criterion at x = 150 m | Maximum smoke layer thickness between curtain and fire (isothermal at 20°C) [m] |
|---|---|---|---|
| 1.00 | No | 2.0 | 2.65 |
| 0.90 | No | 1.9 | 2.85 |
| 0.85 | No | 2.0 | 3.05 |
| 0.80 | No | 1.9 | 3.10 |
| 0.75 | No | 1.9 | 3.30 |
| 0.70 | Yes | - | - |

The loss of confinement is achieved at a velocity of 0.7 m/s, which indicates that the 1 m high curtain provides a 30% reduction in the longitudinal ventilation requirements. Despite the thickening of the smoke layer between the curtain and the fire source as velocity is reduced, blockage is still guaranteed. Finally, the Newman number calculated downstream at 50 m from the fire source seems to confirm that the use of a curtain has no effect on improving stratification downstream of the fire.

## 5    APPLICATION CASE

### 5.1    The operational objectives

The previous results were obtained in an academic context. They suggest that the longitudinal smoke containment is controlled by a combination of smoke curtain height and longitudinal air velocity.

The use of tunnel smoke curtains therefore seems likely to effectively complement the effect of a controlled air flow, in the context of an appropriate management of the mechanical ventilation system.

The effectiveness of this concept is illustrated by the case of a fire in a tunnel equipped with a transverse ventilation system. The aim is to show that a lower extraction capacity coupled with optimal management of longitudinal flows and the installation of smoke curtain is highly effective in terms of smoke stratification and longitudinal containment.

### 5.2    The model features

The tunnel is modelled using the FDS code. It has two traffic lanes. The modelled section is 400 m long. Its cross-section is 10 m x 5 m. An extraction louver is located at the ceiling, in the middle of the section. The smoke curtains are 1 m high, leaving a clearance of 4 m above the pavement. They are distant of 300 m.

The model also includes the representation of stopped vehicles. They may induce local flow disturbances.

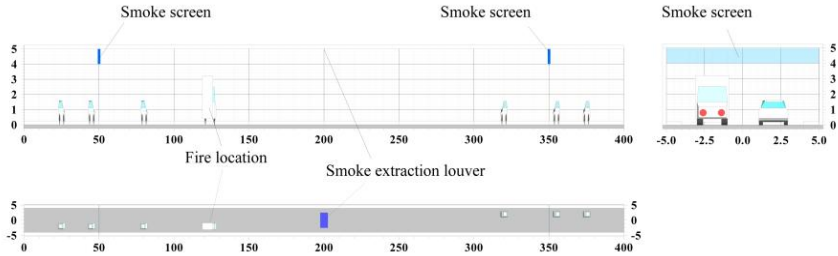

**Figure 4: Main model features.**

The fire affects a van load. Several heat release rates are considered (2 MW, 5 MW and 10 MW).

### 5.3   The simulations

For all calculations, the aim is to ensure that the point of zero longitudinal velocity is maintained below the extract unit (thermal effects make this result imprecise). Therefore, the boundary flows (x = 0 m and x = 400 m) are practically symmetrical.

Two configurations were tested: without and with smoke curtains.

For each of the modelled fire heat release, the extraction rate and velocity at the inlet section are progressively increased. The aim is to identify the aeraulic conditions that will ensure smoke confinement in the vicinity of the smoke curtains.
Situations are characterized for the steady state. This reflects the balance between inertial effects and gravitational stresses.

### 5.4   The performance-based approach

The calculations assume that the flow is controlled at the limits of the section affected by the smoke extension. In practice, this can only be achieved by an active regulation of a part of the tunnel's ventilation system (jet fans or elements of the transverse ventilation system not involved in the smoke extraction).
The most common techniques use thrust, or transverse flow control indexed to analysis of the longitudinal velocity field. They are generally based on PID-type control. They produce satisfactory results in terms of providing the local conditions for the development of natural stratification of hot smoke or controlling the critical velocity (10).

## 5.5 Numerical simulation results and analysis

Longitudinal cross-sections of the temperature field corresponding to smoke confinement in the vicinity of the smoke curtains are shown in the following figures.

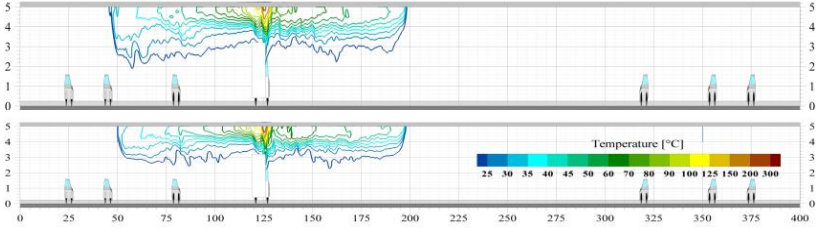

Figure 5: Temperature field longitudinal cross-sections for the smoke confinement, for a 2 MW heat release rate fire (without and with smoke curtains).

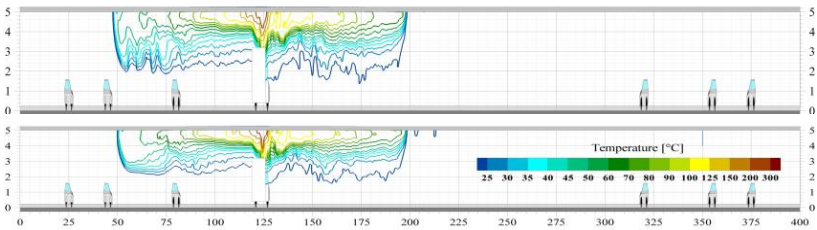

Figure 6: Temperature field longitudinal cross-sections for the smoke confinement, for a 5 MW heat release rate fire (without and with smoke curtains).

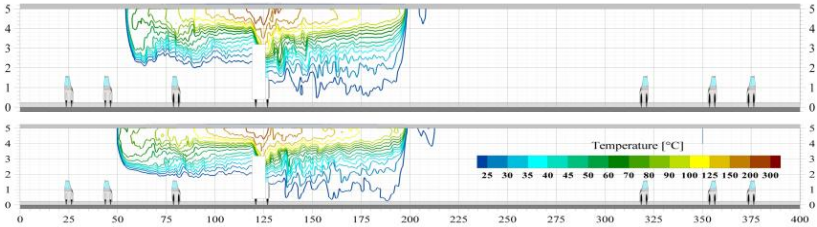

Figure 7: Temperature field longitudinal cross-sections for the smoke confinement, for a 10 MW heat release rate fire (without and with smoke curtains).

The numerical model suggests that, in the absence of a smoke curtain, it is possible to stabilize the backlayering by adjusting the longitudinal air velocity. The aim here is to ensure containment at the position of the curtains.

A second series of calculations is carried out in the presence of curtains. Their height is 1 m. The results show the situation corresponding to the minimum longitudinal velocity required for the confinement. Below this value, some small quantities of smoke can pass below the curtain.

A comparison of the results shows that the state of the smoke is virtually identical in both cases, whatever the fire heat release rate. This observation relates to the state of the stratification upstream and downstream of the fire.

The following figure compares the values for critical and confinement velocities without and with a smoke curtain.

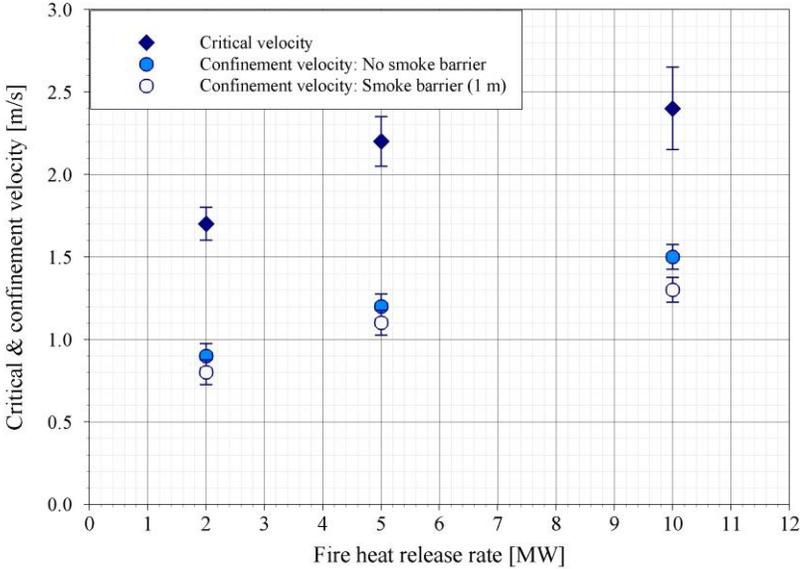

**Figure 8: Values of the critical velocity and the confinement velocity without and with smoke curtains.**

The difference between the critical and the confinement velocity (6) is relatively large (about 1 m/s). It illustrates the fact that the energy required to stop the progression of the backlayering is lower than the energy required to prevent its formation. This result appears natural, since the hot layer loses energy as it progresses, mainly through thermal conducto-convective exchanges with the tunnel walls.

The smoke curtain reduces the confinement velocity. While this contribution appears to be minor (between 10% and 15% of the base velocity for a height of 1 m), an increase in the rate of obstruction should highlight a more marked effect. This aspect will be addressed in future work.

## 6    CONCLUSIONS

The main objective was to study the propagation of a smoke layer in a hybrid ventilation configuration using containment curtains in a tunnel. This case was first studied from an academic point of view in a longitudinally ventilated room equipped with a smoke curtain occupying 1/5 of the tunnel height. After that, these phenomena were studied in an applied case with a smoke extraction louver.

Simulations were carried out on FDS. Firstly, to study the aerodynamics of the smoke curtain, in a case without fire and a ventilation at 1 m/s. Next, a simulation was run, showing that in the case without a curtain, with a 2 MW fire and a longitudinal ventilation of 1 m/s, the smoke propagates 75.5 m upstream of the fire. Next, a similar case was studied, but with a 1 m curtain. The ventilation velocity was progressively reduced until the smoke spills over the curtain. Afterwards, a more applied study was conducted on the use of tunnel curtains in a case where ventilation was provided by extraction louvers.

The following conclusions can be drawn from these simulations:
- Comparing the cases with and without a curtain, the use of a curtain obstructing 1/5th of the tunnel height results in a 30% reduction in ventilation velocity for an identical result in terms of longitudinal propagation.
- In the presence of a smoke curtain, the smoke layer increases in thickness by ~ 20%.
- Stratification downstream of the fire source is unaffected by the reduction of ventilation velocity allowed by the presence of a smoke.

The present work carried out in tunnels suggests that the smoke confinement can be achieved with a significant reduction in the flow rates required for this purpose. These have shown that the use of a smoke curtain reduces the need for longitudinal ventilation to control smoke propagation. Work in progress consists of experimentally reproducing the results obtained numerically on the model.